\documentclass[12 pt,twoside,aps,prd,amsmath,amssymb,
tightenlines,showpacs,showkeys,eqsecnum,pagebackref]{revtex4-1}

\usepackage{graphicx}
\usepackage{mathptmx}
\usepackage{bm}
\newcommand {\pr}{\partial}
\newcommand {\cG}{\cal G}
\newcommand {\cD}{\cal D}
\newcommand {\bg}{\bar \gamma}
\newcommand {\G}{\Gamma}
\newcommand {\bp}{\bar \psi}
\newcommand {\p}{\psi}

\def\myfigure #1#2#3#4
{\begin{figure}[ht]\begin{center}
\includegraphics[width=#2 \textwidth]{#1.eps}
\parbox[t]{#4\textwidth}{\caption{#3}\label{#1}}
\end{center}\end{figure}}

\def \myfigures #1#2#3#4#5#6#7#8
{\begin{figure}[ht]
    \begin{center}
        \includegraphics[width=#2 \textwidth]{#1.eps}
        \hfill
        \includegraphics[width=#6 \textwidth]{#5.eps}
        \parbox[t]{#4\textwidth}{\caption {#3}\label{#1}}
        \hfill
        \parbox[t]{#8\textwidth}{\caption {#7}\label{#5}}
    \end{center}
\end{figure} }

\begin{document}
\baselineskip -24pt
\title{Nonlinear Spinor Fields in Bianchi type-V spacetime}
\author{Bijan Saha}
\affiliation{Laboratory of Information Technologies\\
Joint Institute for Nuclear Research\\
141980 Dubna, Moscow region, Russia} \email{bijan@jinr.ru}
\homepage{http://bijansaha.narod.ru}

\begin{abstract}

A self-consistent system of nonlinear spinor and Bianchi type-V
anisotropic gravitational fields are investigated. It is found that
the presence of nontrivial non-diagonal components of the
energy-momentum tensor of the spinor field imposes some severe
restrictions to the system. As a result two different solutions are
found. In one case the metric functions are similar to each other,
i.e., $a_1 \sim a_2 \sim a_3$ and the spinor mass and spinor field
nonlinearity do not disappear from the system. In this case the
spacetime expands with acceleration in case of a positive
self-coupling constant $\lambda$. A negative $\lambda$ gives rise to
a cyclic or periodical mode of expansion. In the second case the
spinor mass and the spinor field nonlinearity vanish and the
Universe expands linearly with time.
\end{abstract}

\keywords{Spinor field, Bianchi type-V anisotropic cosmological
models}

\pacs{98.80.Cq}

\maketitle

\bigskip

\section{Introduction}

Study of spinor field in cosmology for more than two decades has
firmly established its ability to overcome some longstanding
problems of modern cosmology, such as, the initial singularity,
isotropization of the initially anisotropic spacetime and late time
accelerated mode of expansion of the Universe
\cite{henneaux,ochs,saha1997a,saha1997b,saha2001a,saha2004a,
saha2004b,saha2006c,saha2006e,saha2007,saha2006d,greene,ribas,souza,kremer,
PopPLB,FabIJTP,FabGRG,ELKO,FabJMP,PopPRD,PopGREG}. Especially after
the detection and further experimental reconfirmation of current
cosmic acceleration \cite{riess,perlmutter} the spinor field was
considered as an alternative model for dark energy  in a number of
papers\cite{ribas, saha2006d,saha2006e,saha2007,PopGREG}. Moreover,
it was established that beside describing the different stages of
the evolution of the Universe, spinor field can simulate perfect
fluid and dark energy
\cite{krechet,saha2010a,saha2010b,saha2011,saha2012}.

But some recent studies \cite{sahaIJTP2014,sahaAPSS2015,sahabvi0}
show that the spinor field has more surprise for us. Due to its
specific behavior in curve spacetime the spinor field can
significantly change not only the geometry of spacetime but itself
as well. The existence of nontrivial non-diagonal components of the
energy-momentum tensor plays a vital role in this matter. In
\cite{sahaIJTP2014,sahaAPSS2015} it was shown that depending on the
type restriction imposed on the non-diagonal components of the
energy-momentum tensor, the initially Bianchi type-I evolve into a
LRS Bianchi type-I spacetime or FRW one from the very beginning,
whereas the model may describe a general Bianchi type-I spacetime
but in that case the spinor field becomes massless and linear. The
same thing happens for a Bianchi type-$VI_0$ spacetime, i.e., the
geometry of Bianchi type-$VI_0$ spacetime does not allow the
existence of a massive and nonlinear spinor field in some particular
cases \cite{sahabvi0}.

A Bianchi type-V model describes an anisotropic but homogeneous
Universe. This model was studied by several authors
\cite{saha2004b,GCBVI2010,svBVI,Hoogen,Socorro,Weaver1}, specially
due to the existence of magnetic fields in galaxies which was proved
by a number of astrophysical observations. Whereas, some dark energy
model within the scope of a BV cosmology was studied in
\cite{sahabv,sahaECAADE,Yadav2011}.

The purpose of this paper is to investigate the role of Bianchi
type-V geometry on the evolution of the spinor field and vice versa.

\section{Basic equation}

Let us consider the case when the anisotropic spacetime is filled
with nonlinear spinor field. The corresponding action can be given
by
\begin{equation}
{\cal S}(g; \psi, \bp) = \int\, L \sqrt{-g} d\Omega \label{action}
\end{equation}
with
\begin{equation}
L= L_{\rm g} + L_{\rm sp}. \label{lag}
\end{equation}
Here $L_{\rm g}$ corresponds to the gravitational field
\begin{equation}
L_{\rm g} = \frac{R}{2\kappa}, \label{lgrav}
\end{equation}
where $R$ is the scalar curvature, $\kappa = 8 \pi G$, with G being
Newton's gravitational constant and $L_{\rm sp}$ is the spinor field
Lagrangian.

\subsection{Gravitational field}

The gravitational field in our case is given by a Bianchi type-V
anisotropic spacetime:

\begin{equation}
ds^2 = dt^2 - a_1^2 e^{2mx_3} dx_1^2 - a_2^2 e^{2mx_3} dx_2^2 -
a_3^2 dx_3^2, \label{bv}
\end{equation}
with $a_1,\,a_2$ and $a_3$ being the functions of time only and $m$
is some arbitrary constants.

The nontrivial Christoffel symbols for \eqref{bv} are
\begin{eqnarray}
\G_{01}^{1} &=& \frac{\dot{a_1}}{a_1},\quad \G_{02}^{2} =
\frac{\dot{a_2}}{a_2},\quad
\G_{03}^{3} = \frac{\dot{a_3}}{a_3}, \nonumber\\
\G_{11}^{0} &=& a_1 \dot{a_1} e^{2mx_3},\quad \G_{22}^{0} = a_2
\dot{a_2} e^{2mx_3},\quad
\G_{33}^{0} = a_3 \dot{a_3},\label{Chrysvi}\\
\G_{31}^{1} &=& m,\quad \G_{32}^{2} = m,\quad \G_{11}^{3} = -\frac{m
a_1^2}{a_3^2} e^{2mx_3},\quad \G_{22}^{3} = - \frac{m a_2^2}{a_3^2}
e^{2mx_3}. \nonumber
\end{eqnarray}

The nonzero components of the Einstein tensor corresponding to the
metric \eqref{bv} are
\begin{subequations}
\label{ET}
\begin{eqnarray}
G_1^1 &=&  -\frac{\ddot a_2}{a_2} - \frac{\ddot a_3}{a_3} -
\frac{\dot a_2}{a_2}\frac{\dot a_3}{a_3} + \frac{m^2}{a_3^2}, \label{ET11}\\
G_2^2 &=&  -\frac{\ddot a_3}{a_3} - \frac{\ddot a_1}{a_1} -
\frac{\dot a_3}{a_3}\frac{\dot a_1}{a_1} + \frac{m^2}{a_3^2}, \label{ET22}\\
G_3^3 &=&  -\frac{\ddot a_1}{a_1} - \frac{\ddot a_2}{a_2} -
\frac{\dot a_1}{a_1}\frac{\dot a_2}{a_2} + \frac{m^2}{a_3^2}, \label{ET33}\\
G_0^0 &=&  -\frac{\dot a_1}{a_1}\frac{\dot a_2}{a_2} - \frac{\dot
a_2}{a_2}\frac{\dot a_3}{a_3} - \frac{\dot a_3}{a_3}\frac{\dot
a_1}{a_1} + \frac{3m^2}{a_3^2}, \label{ET00}\\
G_3^0 &=& m \Bigl(\frac{\dot a_1}{a_1} +
 \frac{\dot a_2}{a_2}  - 2  \frac{\dot a_3}{a_3}\Bigr). \label{ET03}
\end{eqnarray}
\end{subequations}

\subsection{Spinor field}

Keeping in mind the symmetry between $\p$ and $\bp$ \cite{kibble} we
choose the spinor field Lagrangian as \cite{saha2001a}:
\begin{equation}
L_{\rm sp} = \frac{\imath}{2} \biggl[\bp \gamma^{\mu} \nabla_{\mu}
\psi- \nabla_{\mu} \bar \psi \gamma^{\mu} \psi \biggr] - m_{\rm sp}
\bp \psi - F, \label{lspin}
\end{equation}
where the nonlinear term $F$ describes the self-interaction of a
spinor field and can be presented as some arbitrary functions of
invariants generated from the real bilinear forms of a spinor field.
On account of Fierz identity we choose the nonlinear term $F = F(K)$
with $K$ taking one of the following expressions
$\{I,\,J,\,I+J,\,I-J\}$, where $I = S^2 = \left(\bar \psi
\psi\right)^2$ and $J = P^2 = \left(\imath \bar \psi \gamma^5
\psi\right)^2$. As it was shown in \cite{saha2001a}, such a
nonlinear term describes the nonlinearity in its most general form.
Here $\nabla_\mu$ is the covariant derivative of spinor field:
\begin{equation}
\nabla_\mu \psi = \frac{\partial \psi}{\partial x^\mu} -\G_\mu \psi,
\quad \nabla_\mu \bp = \frac{\partial \bp}{\partial x^\mu} + \bp
\G_\mu, \label{covder}
\end{equation}
with $\G_\mu$ being the spinor affine connection. In \eqref{lspin}
$\gamma$'s are the Dirac matrices in curve spacetime and obey the
following algebra
\begin{equation}
\gamma^\mu \gamma^\nu + \gamma^\nu \gamma^\mu = 2 g^{\mu\nu}
\label{al}
\end{equation}
and are connected with the flat spacetime Dirac matrices $\bg$ in
the following way
\begin{equation}
 g_{\mu \nu} (x)= e_{\mu}^{a}(x) e_{\nu}^{b}(x) \eta_{ab},
\quad \gamma_\mu(x)= e_{\mu}^{a}(x) \bg_a \label{dg}
\end{equation}
where $e_{\mu}^{a}$ is a set of tetrad 4-vectors.

For the metric \eqref{bv} we choose the tetrad as follows:

\begin{equation}
e_0^{(0)} = 1, \quad e_1^{(1)} = a_1 e^{mx_3}, \quad e_2^{(2)} = a_2
e^{mx_3}, \quad e_3^{(3)} = a_3. \label{tetradvi}
\end{equation}

The Dirac matrices $\gamma^\mu(x)$ of Bianchi type-V spacetime are
connected with those of Minkowski one as follows:
$$ \gamma^0=\bg^0,\quad \gamma^1 = \frac{ e^{-m x_3}}{a_1} \bg^1,
\quad \gamma^2= \frac{ e^{-m x_3}}{a_2}\bg^2,\quad \gamma^3 = \frac{
1}{a_3}\bg^3$$

$$\gamma^5 = - \imath \sqrt{-g}
\gamma^0\gamma^1\gamma^2\gamma^3 = - \imath \bg^0\bg^1\bg^2\bg^3 =
\bg^5
$$
with
\begin{eqnarray}
\bg^0\,&=&\,\left(\begin{array}{cccc}1&0&0&0\\0&1&0&0\\0&0&-1&0\\0&0&0&-1\end{array}\right),
\quad
\bg^1\,=\,\left(\begin{array}{cccc}0&0&0&1\\0&0&1&0\\0&-1&0&0\\-1&0&0&0\end{array}\right),
\quad
\bg^2\,=\,\left(\begin{array}{cccc}0&0&0&-\imath\\0&0&\imath&0\\0&\imath&0&0\\-\imath&0&0&0\end{array}\right),\\
\bg^3\,&=&\,\left(\begin{array}{cccc}0&0&1&0\\0&0&0&-1\\-1&0&0&0\\0&1&0&0\end{array}\right),
\quad \gamma^5 = \bg^5 =
\left(\begin{array}{cccc}0&0&-1&0\\0&0&0&-1\\-1&0&0&0\\0&-1&0&0\end{array}\right).\nonumber
\end{eqnarray}

The spinor affine connection matrices $\G_\mu (x)$ are uniquely
determined up to an additive multiple of the unit matrix by the
equation
\begin{equation}
\frac{\pr \gamma_\nu}{\pr x^\mu} - \G_{\nu\mu}^{\rho}\gamma_\rho -
\G_\mu \gamma_\nu + \gamma_\nu \G_\mu = 0, \label{afsp}
\end{equation}
with the solution
\begin{equation}
\Gamma_\mu = \frac{1}{4} \bg_{a} \gamma^\nu \partial_\mu e^{(a)}_\nu
- \frac{1}{4} \gamma_\rho \gamma^\nu \Gamma^{\rho}_{\mu\nu}.
\label{sfc}
\end{equation}

For the Bianchi type-V metric \eqref{sfc} one finds the following
expressions for spinor affine connections:
\begin{subequations}
\label{sac123}
\begin{eqnarray}
\G_0 &=& 0, \label{sac0}\\  \G_1 &=& \frac{1}{2}\Bigl(\dot a_1
\bg^1\bg^0 + m\frac{a_1}{a_3} \bg^1\bg^3\Bigr) e^{mx_3},
\label{sac1}\\  \G_2 &=& \frac{1}{2}\Bigl(\dot a_2 \bg^2\bg^0 +
m\frac{a_2}{a_3} \bg^2\bg^3\Bigr) e^{mx_3}, \label{sac2}\\  \G_3 &=&
\frac{\dot a_3}{2} \bg^3 \bg^0. \label{sac3}
\end{eqnarray}
\end{subequations}

\subsection{Field equations}

Variation of \eqref{action} with respect to the metric function
$g_{\mu \nu}$ gives the Einstein field equation
\begin{equation}
G_\mu^\nu = R_\mu^\nu - \frac{1}{2} \delta_\mu^\nu R = -\kappa
T_\mu^\nu, \label{EEg}
\end{equation}
where $R_\mu^\nu$ and $R$ are the Ricci tensor and Ricci scalar,
respectively. Here $T_\mu^\nu$ is the energy-momentum tensor of the
spinor field.

Varying \eqref{lspin} with respect to $\bp (\psi)$ one finds the
spinor field equations:
\begin{subequations}
\label{speq}
\begin{eqnarray}
\imath\gamma^\mu \nabla_\mu \psi - m_{\rm sp} \psi - {\cD} \psi -
 \imath {\cG} \gamma^5 \psi &=&0, \label{speq1} \\
\imath \nabla_\mu \bp \gamma^\mu +  m_{\rm sp} \bp + {\cD}\bp +
\imath {\cG} \bp \gamma^5 &=& 0, \label{speq2}
\end{eqnarray}
\end{subequations}
where we denote ${\cD} = 2 S F_K K_I$ and ${\cG} = 2 P F_K K_J$,
with $F_K = dF/dK$, $K_I = dK/dI$ and $K_J = dK/dJ.$

\subsection{Energy-momentum tensor of the spinor field}

The energy-momentum tensor of the spinor field is given by
\begin{equation}
T_{\mu}^{\rho}=\frac{\imath}{4} g^{\rho\nu} \biggl(\bp \gamma_\mu
\nabla_\nu \psi + \bp \gamma_\nu \nabla_\mu \psi - \nabla_\mu \bar
\psi \gamma_\nu \psi - \nabla_\nu \bp \gamma_\mu \psi \biggr) \,-
\delta_{\mu}^{\rho} L_{\rm sp}. \label{temsp}
\end{equation}

Then in view of \eqref{covder} the energy-momentum tensor of the
spinor field can be written as
\begin{eqnarray}
T_{\mu}^{\,\,\,\rho}&=&\frac{\imath}{4} g^{\rho\nu} \bigl(\bp
\gamma_\mu
\partial_\nu \psi + \bp \gamma_\nu \partial_\mu \psi -
\partial_\mu \bar \psi \gamma_\nu \psi - \partial_\nu \bp \gamma_\mu
\psi \bigr)\nonumber\\
& - &\frac{\imath}{4} g^{\rho\nu} \bp \bigl(\gamma_\mu \G_\nu +
\G_\nu \gamma_\mu + \gamma_\nu \G_\mu + \G_\mu \gamma_\nu\bigr)\psi
 \,- \delta_{\mu}^{\rho} \bigl(2 K F_K - F(K)\bigr). \label{temsp0}
\end{eqnarray}
In \eqref{temsp0} we used the fact that in view of \eqref{speq} the
spinor field Lagrangian can be rewritten as
\begin{eqnarray}
L_{\rm sp} = 2 (I F_I + J F_J) - F = 2 K F_K - F(K). \label{lspin01}
\end{eqnarray}
In what follows we consider the case when the spinor field depends
on $t$ only, i.e. $\psi = \psi (t)$. As it is seen from
\eqref{temsp0}, in case if for a given metric $\G_\mu$'s are
different, there arise nontrivial non-diagonal components of the
energy-momentum tensor. Thus, for the case in hand, after a little
manipulations from \eqref{temsp0} we find
\begin{subequations}
\label{Ttot}
\begin{eqnarray}
T_0^0 & = & m_{\rm sp} S + F(K), \label{emt00}\\
T_1^1 &=& T_2^2 = T_3^3 =  F(K) - 2 K F_K, \label{emtii}\\
T_1^0 &=& -\frac{\imath}{4} m \frac{a_1}{a_3} e^{m x_3}\, \bp \bg^3
\bg^1 \bg^0 \psi
= -\frac{1}{4} m \frac{a_1}{a_3} e^{m x_3}\, A^2 , \label{emt01} \\
T_2^0 &=&\frac{\imath}{4} m \frac{a_2}{a_3} e^{m x_3}\, \bp \bg^2
\bg^3 \bg^0 \psi
= \frac{1}{4} m \frac{a_2}{a_3} e^{m x_3}\,A^1, \label{emt02} \\
T_3^0 &=& 0, \label{emt03} \\
T_2^1 &=& \frac{\imath}{4} \frac{a_2}{a_1} \biggl(\frac{\dot
a_1}{a_1} - \frac{\dot a_2}{a_2}\biggr) \bp \bg^1 \bg^2 \bg^0 \psi =
\frac{1}{4} \frac{a_2}{a_1}\biggl(\frac{\dot a_1}{a_1} - \frac{\dot
a_2}{a_2}\biggr) A^3, \label{emt12}\\
T_3^1 &=&\frac{\imath}{4} \frac{a_3}{a_1} e^{-m x_3}
\biggl(\frac{\dot a_3}{a_3} - \frac{\dot a_1}{a_1}\biggr) \bp \bg^3
\bg^1 \bg^0 \psi = \frac{1}{4} \frac{a_3}{a_1} e^{-m x_3}
\biggl(\frac{\dot a_3}{a_3} - \frac{\dot a_1}{a_1}\biggr) A^2 \label{emt13}\\
T_3^2 &=&\frac{\imath}{4} \frac{a_3}{a_2} e^{-m x_3}
\biggl(\frac{\dot a_2}{a_2} - \frac{\dot a_3}{a_3}\biggr) \bp \bg^2
\bg^3 \bg^0 \psi = \frac{1}{4} \frac{a_3}{a_2} e^{-m x_3}
\biggl(\frac{\dot a_2}{a_2} - \frac{\dot a_3}{a_3}\biggr)A^1.
\label{emt23}
\end{eqnarray}
\end{subequations}

It can be shown that bilinear spinor forms the obey the following
system of equations:
\begin{subequations}
\label{inv}
\begin{eqnarray}
\dot S_0  +  {\cG} A_{0}^{0} &=& 0, \label{S0} \\
\dot P_0  -  \Phi A_{0}^{0} &=& 0, \label{P0}\\
\dot A_{0}^{0} +\frac{2m}{a_3} A_{0}^{3} +  \Phi P_0 -  {\cG}
S_0 &=& 0, \label{A00}\\
\dot A_{0}^{3} +\frac{2m}{a_3} A_{0}^{0} &=& 0, \label{A03}\\
\dot v_{0}^{0} + \frac{2m}{a_3} v_{0}^{3} &=& 0,\label{v00} \\
\dot v_{0}^{3} + \frac{2m}{a_3} v_{0}^{0} +
\Phi Q_{0}^{30} +  {\cG} Q_{0}^{21} &=& 0,\label{v03}\\
\dot Q_{0}^{30}  -  \Phi v_{0}^{3} &=& 0,\label{Q030} \\
\dot Q_{0}^{21}  -  {\cG} v_{0}^{3} &=& 0, \label{Q021}
\end{eqnarray}
\end{subequations}
where we denote ${\cal X}_0 = V {\cal X}$ and $\Phi = m_{\rm sp} +
{\cD}$. Here we also define the Volume scale
\begin{equation}
V = a_1 a_2 a_3. \label{VDef}
\end{equation}

In \eqref{inv}, $S = \bar \psi \psi$ is a scalar, $
 P =  \imath \bar \psi \gamma^5 \psi$ is a pseudoscalar,
$ v^\mu = \bar \psi \gamma^\mu \psi$ - vector, $ A^\mu = \bar \psi
\gamma^5 \gamma^\mu \psi$ - pseudovector, and  $Q^{\mu\nu} =\bar
\psi \sigma^{\mu\nu} \psi$ is antisymmetric tensor.

Combining these equations together and taking the first integral one
gets the following relations:
\begin{subequations}
\label{inv0}
\begin{eqnarray}
(S_{0})^{2} + (P_{0})^{2} + (A_{0}^{0})^{2} - (A_{0}^{3})^{2} &=&
C_1 = {\rm Const}, \label{inv01}\\
(Q_{0}^{30})^{2} + (Q_{0}^{21})^{2} + (v_{0}^{3})^{2} -
(v_{0}^{0})^{2} &=& C_2 = {\rm Const.} \label{inv02}
\end{eqnarray}
\end{subequations}

\section{Solution to the field equations}

In this section we solve the field equations. First of all we write
the expression for $K$. From \eqref{S0} and \eqref{P0} it can be
shown that \cite{sahabvi0}
\begin{equation}
K = \frac{V_0^2}{V^2}. \label{Ksol}
\end{equation}

Note that the expression \eqref{Ksol} holds for $K = I$ for both
massive and massless spinor field, whereas, for $K =
\{J,\,I+J,\,I-J\}$  it holds only for massless spinor field
\cite{sahaAPSS2015,sahabvi0}.

Let us begin with the spinor field equations. In view of
\eqref{covder} and \eqref{sac123} the spinor field equation
\eqref{speq} takes the form

\begin{subequations}
\label{SF1}
\begin{eqnarray}
\imath \bg^0 \bigl(\dot \psi + \frac{1}{2}\frac{\dot V}{V}
\psi\bigr) - m_{\rm sp} \psi +\frac{m}{a_3} \bg^3 \psi - {\cD}
\psi -  \imath {\cG} \bg^5 \psi &=&0, \label{speq1p}\\
\imath \bigl(\dot \bp + \frac{1}{2}\frac{\dot V}{V} \bp\bigr)\bg^0 +
m_{\rm sp} \bp +\frac{m}{a_3}\bp \bg^3   + {\cD}  \bp + \imath
{\cG}\bp \bg^5 &=& 0. \label{speq2p}
\end{eqnarray}
\end{subequations}

As we have already mentioned, $\psi$ is a function of $t$ only. We
consider the 4-component spinor field given by
\begin{eqnarray}
\psi = \left(\begin{array}{c} \psi_1\\ \psi_2\\ \psi_3 \\
\psi_4\end{array}\right). \label{psi}
\end{eqnarray}
Denoting $\phi_i =\sqrt{V} \psi_i$ and $\bar X_0 = -m X_0^{1/3}$
from \eqref{SF1} for the spinor field we find we find
\begin{subequations}
\label{speq1pfg}
\begin{eqnarray}
\dot \phi_1 + \imath\, {\Phi} \phi_1 + \Bigl[\imath\, \frac{\bar X_0}{V^{1/3}} + {\cG}\Bigr] \phi_3 &=& 0, \label{ph1}\\
\dot \phi_2 + \imath\, {\Phi} \phi_2 - \Bigl[\imath\, \frac{\bar X_0}{V^{1/3}} - {\cG}\Bigr] \phi_4 &=& 0, \label{ph2}\\
\dot \phi_3 - \imath\, {\Phi} \phi_3 + \Bigl[\imath\, \frac{\bar X_0}{V^{1/3}} -  {\cG}\Bigr] \phi_1 &=& 0, \label{ph3}\\
\dot \phi_4 - \imath\, {\Phi} \phi_4 - \Bigl[\imath\, \frac{\bar
X_0}{V^{1/3}} + {\cG} \Bigr] \phi_2&=& 0. \label{ph4}
\end{eqnarray}
\end{subequations}
The foregoing system of equations can be written in the form:
\begin{equation}
\dot \phi = A \phi, \label{phi}
\end{equation}
with $\phi = {\rm
col}\left(\phi_1,\,\phi_2,\,\phi_3,\,\phi_4\right)$ and
\begin{equation}
A = \left(\begin{array}{cccc}-\imath\, \Phi &0 & -\imath\, {\cal Y}
- {\cG}& 0 \\ 0&-\imath\, \Phi& 0 &\imath\, {\cal Y} -
{\cG}\\-\imath\, {\cal Y} + {\cG}&0&\imath\, \Phi&0\\0&\imath\,
{\cal Y} + {\cG} & 0 &\imath\, \Phi \end{array}\right). \label{AMat}
\end{equation}
where we denote ${\cal Y} = \frac{\bar X_0}{V^{1/3}}$. It can be
easily found that
\begin{equation}
{\rm det} A = \left(\Phi^2 + {\cal Y}^2 +{\cG}^2\right)^2.
\label{detA}
\end{equation}

The solution to the equation \eqref{phi} can be written in the form
\begin{equation}
\phi(t) = {\rm T exp}\Bigl(-\int_t^{t_1}  A_1 (\tau) d \tau\Bigr)
\phi (t_1), \label{phi1}
\end{equation}
where
\begin{equation}
A_1 = \left(\begin{array}{cccc}-\imath\, {\cD} &0 & -\imath\, {\cal
Y} - {\cG}& 0 \\ 0&-\imath\, {\cD}& 0 &\imath\, {\cal Y} -
{\cG}\\-\imath\, {\cal Y} + {\cG}&0&\imath\, {\cD}&0\\0&\imath\,
{\cal Y} + {\cG} & 0 &\imath\, {\cD} \end{array}\right).
\label{AMat1}
\end{equation}
and $\phi (t_1)$ is the solution at $t = t_1$. As we have already
shown, $K = V_0^2/V^2$ for $K = \{J,\,I+J,\,I-J\}$ with trivial
spinor-mass and $K = V_0^2/V^2$ for $K=I$ for any spinor-mass. Since
our Universe is expanding, the quantities ${\cD}$, ${\cal Y}$ and
${\cG}$ become trivial at large $t$. Hence in case of  $K = I$ with
non-trivial spinor-mass one can assume $\phi (t_1) = {\rm
col}\left(e^{-\imath m_{\rm sp} t_1},\,e^{-\imath m_{\rm sp}
t_1},\,e^{\imath m_{\rm sp} t_1},\,e^{\imath m_{\rm sp}
t_1}\right)$, whereas for other cases with trivial spinor-mass we
have $\phi (t_1) = {\rm
col}\left(\phi_{1}^{0},\,\phi_{2}^{0},\,\phi_{3}^{0},\,\phi_{4}^{0}\right)$
with $\phi_i^0$ being some constants. Here we have used the fact
that $\Phi = m_{\rm sp} + {\cD}.$ The other way to solve the system
\eqref{speq1pfg} is given in \cite{saha2004b}.

Now let us consider the Einstein field equations. In view of
\eqref{ET} and \eqref{Ttot} with find the following system of
Einstein Equations

\begin{subequations}
\label{EE}
\begin{eqnarray}
\frac{\ddot a_2}{a_2} + \frac{\ddot a_3}{a_3} +
\frac{\dot a_2}{a_2}\frac{\dot a_3}{a_3} - \frac{m^2}{a_3^2} &=& \kappa\bigl(F(K) - 2 K F_K\bigr), \label{EE11}\\
\frac{\ddot a_3}{a_3} + \frac{\ddot a_1}{a_1} +
\frac{\dot a_3}{a_3}\frac{\dot a_1}{a_1} - \frac{m^2}{a_3^2} &=& \kappa\bigl(F(K) - 2 K F_K\bigr), \label{EE22}\\
\frac{\ddot a_1}{a_1} + \frac{\ddot a_2}{a_2} +
\frac{\dot a_1}{a_1}\frac{\dot a_2}{a_2} - \frac{m^2}{a_3^2} &=& \kappa\bigl(F(K) - 2 K F_K\bigr), \label{EE33}\\
\frac{\dot a_1}{a_1}\frac{\dot a_2}{a_2} + \frac{\dot
a_2}{a_2}\frac{\dot a_3}{a_3} + \frac{\dot a_3}{a_3}\frac{\dot
a_1}{a_1} - \frac{3m^2}{a_3^2} &=&   \kappa\bigl(m_{\rm sp} S +
F(K)\bigr), \label{EE00}
\end{eqnarray}
\end{subequations}
with the additional constraints
\begin{subequations}
\label{AC}
\begin{eqnarray}
\frac{\dot a_1}{a_1} +  \frac{\dot a_2}{a_2} - 2 \frac{\dot
a_3}{a_3}   &=& 0, \label{EE03}\\
\frac{1}{4} m \frac{a_1}{a_3} e^{m x_3}\, A^2 &=& 0, \label{AC01} \\
\frac{1}{4} m \frac{a_2}{a_3} e^{m x_3}\,A^1 &=& 0, \label{AC02} \\
\frac{1}{4} \frac{a_2}{a_1} \biggl(\frac{\dot a_1}{a_1} - \frac{\dot
a_2}{a_2}\biggr) A^3
 &=& 0, \label{AC12}\\
\frac{1}{4} \frac{a_3}{a_1} e^{-m x_3}
\biggl(\frac{\dot a_3}{a_3} - \frac{\dot a_1}{a_1}\biggr) A^2 &=& 0, \label{AC13}\\
\frac{1}{4} \frac{a_3}{a_2} e^{-m x_3} \biggl(\frac{\dot a_2}{a_2} -
\frac{\dot a_3}{a_3}\biggr)A^1 &=& 0. \label{AC23}
\end{eqnarray}
\end{subequations}

From \eqref{EE03} we have the following relations between the metric
functions:

\begin{equation}
a_1 a_2 = X_2  a_3^2, \quad X_2 = {\rm const.}  \label{a123}
\end{equation}

On the other hand from \eqref{AC01} and \eqref{AC02} one dully finds
\begin{equation}
A^2 = 0, \quad {\rm and} \quad A^1 = 0. \label{A12}
\end{equation}
In view of \eqref{A12} the relations \eqref{AC13} and \eqref{AC23}
fulfill even without imposing restrictions on the metric functions.
From \eqref{AC12} one finds there are two distinct possibilities:
(i) impose restriction on the metric function; (ii) impose it on the
spinor field:
\begin{subequations}
\label{A0A3}
\begin{eqnarray}
\frac{\dot a_1}{a_1} - \frac{\dot a_2}{a_2} &=& 0, \label{A0A3a}\\
A^3 &=& 0. \label{A0A3b}
\end{eqnarray}
\end{subequations}

{\bf Case I} Imposing the restriction on the metric function
\eqref{A0A3a} we find the following relations between $a_1$ and
$a_2$:
\begin{equation}
a_2 =X_1  a_1, \quad X_1 = {\rm const.}  \label{a12n}
\end{equation}
Then in view of \eqref{VDef} from \eqref{a123} and \eqref{a12n} for
the metric functions we finally find
\begin{eqnarray}
a_1 = \frac{X_2^{1/6}}{\sqrt{X_1}} V^{\frac{1}{3}}, \quad a_2 =
X_2^{1/6}\sqrt{X_1} V^{\frac{1}{3}},\quad a_3 = X_2^{-1/3}
V^{\frac{1}{3}}. \label{Metf}
\end{eqnarray}

Thus we see that the spinor field nonlinearity leads to $a_1 \sim
a_2 \sim a_3$ from the very beginning, if restriction is imposed on
the metric functions. Similar result was found for Bianchi type-I
spacetime in \cite{sahaAPSS2015}. It should be noted that thanks to
non-diagonal components of the energy-momentum tensor of the spinor
field in this case we could find the expressions for the metric
functions without additional condition such as proportionality of
shear and expansion.

Thus the expressions for the metric functions are obtained in terms
of volume scale $V$. The equation for $V$ can be derived from the
Einstein Equation \eqref{ET} which after some manipulations looks
\begin{equation}
\ddot V = 6m^2 X_2^{2/3} V^{1/3} + \frac{3 \kappa}{2} \bigl[m_{\rm
sp} S + 2 \bigl(F(K) - K F_K\bigr)\bigr] V. \label{Vdefein}
\end{equation}
Further giving the concrete form of $F(K)$ one can draw the picture
of evolution of the Universe. As it was shown earlier, $K =
V_0^2/V^2$ for all cases with a trivial spinor-mass, whereas it is
so for $K = I = S^2$ even with a non-zero spinor-mass. Hence we
consider the case with $K = I$, setting $F = \lambda I^n$. Then
inserting it into \eqref{Vdefein} we find the following equation

\begin{equation}
\ddot V = \Phi_1(V), \quad \Phi_1(V) \equiv 6m^2 X_2^{2/3} V^{1/3} +
\frac{3 \kappa}{2} \bigl[m_{\rm sp} V_0 + 2 \lambda (1 - n) V_0^{2n}
V^{1 - 2 n}\bigr], \label{Vdefv}
\end{equation}
with the first integral
\begin{equation}
\dot V = \Phi_2(V), \quad \Phi_2(V) \equiv \sqrt{9m^2 X_2^{2/3}
V^{4/3} + 3 \kappa \bigl[m_{\rm sp} V_0 V + \lambda V_0^{2n} V^{2 (1
- n)}\bigr] + V_c}, \quad V_c = {\rm const.} \label{firstint}
\end{equation}
The solution to the foregoing equation can be formally presented as
a quadrature as
\begin{equation}
\int \frac{dV}{\sqrt{9m^2 X_2^{2/3} V^{4/3} + 3 \kappa \bigl[m_{\rm
sp} V_0 V +  \lambda V_0^{2n} V^{2 (1 - n)}\bigr] + V_c}} = t + t_0,
\quad t_0 = {\rm const.} \label{quad}
\end{equation}

The equation \eqref{Vdefv} can be solved numerically. In doing so we
have to give concrete value of problem parameters $m,\,X_2,\,
\kappa,\,m_{\rm sp},\,\lambda,\,V_0,\,n,\,V_c$ and initial value of
$V = V(0)$, whereas the value of $\dot V (0)$ should be calculated
from \eqref{firstint}, which being a square root imposes some
natural restriction on the choice of problem parameters and initial
value.

To determine the character of the evolution, let us first study the
asymptotic behavior of the equation \eqref{Vdefv}. It should be
recall that we have $K = V_0^2 / V^2$. Since all the physical
quantities constructed from the spinor fields as well as the
invariants of gravitational fields are inverse function of $V$ of
some degree, it can be concluded that at any spacetime point where
the volume scale becomes zero, it is a singular point
\cite{saha2001a}. So we assume at the beginning $V$ was small but
non-zero. Then from \eqref{Vdefv} we see that at $t \to 0$ the term
$m_{\rm sp} V_0$ prevails if $1 - 2 n > 0$, i.e., $n < 1/2$. In case
of $n = 1/2$ this term can be added to the mass term. And finally
for $n > 1/2$ the nonlinear term becomes predominant. Recalling that
we are considering an expanding Universe, at $t \to \infty$ the
volume scale should be quite large. In that case the term $6m^2
X_2^{2/3} V^{1/3}$ prevails if  $1 - 2 n < 1/3$, i.e., $n > 1/3$,
while for $n < 1/3$ the nonlinear term prevails at $t \to \infty$.
Since we are mainly interested in the nonlinearity that gives rise
to late time acceleration, it is safe to consider $n < 1/3$
including the negative value for $n$.

To determine the character of expansion at large time let us
introduce the deceleration parameter, which we determine as

\begin{equation}
q = - \frac{V \ddot V}{\dot V^2}, \label{decel}
\end{equation}
which in view of \eqref{Vdefv} and \eqref{firstint} can be written
as

\begin{equation}
q = - \frac{V \Phi_1(V)}{\Phi_2^2 (V)} = - \frac{6m^2 X_2^{2/3}
V^{4/3} + \frac{3 \kappa}{2} \bigl[m_{\rm sp} V_0 V + 2 \lambda (1 -
n) V_0^{2n} V^{2(1 -  n)}\bigr]}{9m^2 X_2^{2/3} V^{4/3} + 3 \kappa
\bigl[m_{\rm sp} V_0 V + \lambda V_0^{2n} V^{2 (1 - n)}\bigr] +
V_c}. \label{decel1}
\end{equation}
Now let us see what happens to $q$ at large $t$.

If we consider the case with $n < 1/3$ when the nonlinear term
decisively predominates, at large $t$ we can rewrite \eqref{decel1}
as
\begin{eqnarray}
q \bigl|_{t \to \infty} \approx  - \frac{3 \kappa \lambda (1 - n)
V_0^{2n} V^{2(1 -  n)}}{3 \kappa \lambda V_0^{2n} V^{2 (1 - n)}} = -
(1 - n) < 0, \quad {\rm for} \quad n < 1/3, \label{decel2}
\end{eqnarray}
whereas, if we consider $n > 1/3$ with predominant $6m^2 X_2^{2/3}
V^{1/3}$ as large $t$, for $q$ in this case we find

\begin{eqnarray}
q \bigl|_{t \to \infty} \approx  - \frac{6m^2 X_2^{2/3}
V^{4/3}}{9m^2 X_2^{2/3} V^{4/3}} = - \frac{2}{3} < 0, \quad {\rm
for} \quad n > 1/3. \label{decel3}
\end{eqnarray}

Thus we conclude that the models gives rise to an accelerated mode
of expansion.

\begin{figure}[ht]
\centering
\includegraphics[height=70mm]{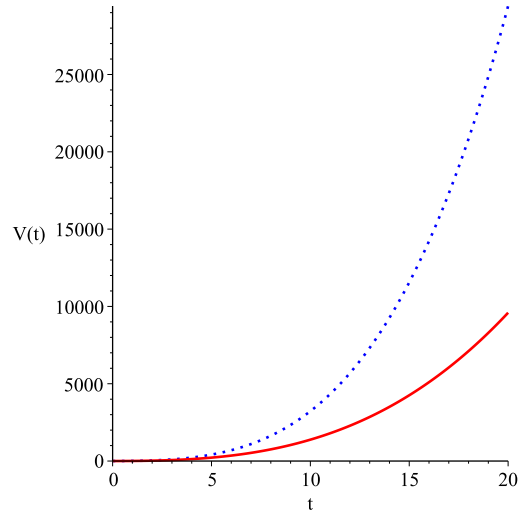} \\
\vskip 1 cm \caption{Evolution of the volume scale $V$ of the
Universe filled with massive spinor field. Here we consider the
following values of problem parameters $m =1,\,X_2 = 1,\, \kappa =
1,\,m_{\rm sp}= 1,\,\lambda = 1,\,V_0 = 1,\,V_c = 10$ and $V(0) =
0.9$.  The blue dotted line stands for the value of $n = 1/4$,
whereas the solid red line corresponds to $n = 2$.}
\label{Vblue025red2}.
\end{figure}

\begin{figure}[ht]
\centering
\includegraphics[height=70mm]{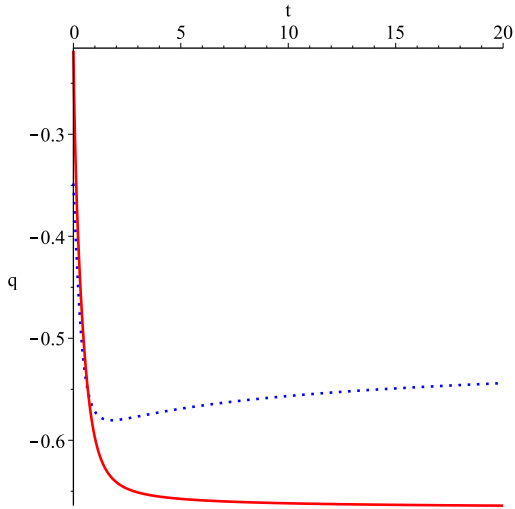} \\
\vskip 1 cm \caption{Evolution of the deceleration parameter $q$.
Here we consider the following values of problem parameters $m
=1,\,X_2 = 1,\, \kappa = 1,\,m_{\rm sp}= 1,\,\lambda = 1,\,V_0 =
1,\,V_c = 10$ and $V(0) = 0.9$.  The blue dotted line stands for the
value of $n = 1/4$, whereas the solid red line corresponds to $n =
2$.} \label{qblue025red2}.
\end{figure}

In Figs. \ref{Vblue025red2} and \ref{qblue025red2} we have plotted
the evolution of volume scale $V(t)$ and deceleration parameter $q$,
respectively. In doing so we have considered the following problem
parameters:  $m =1,\,X_2 = 1,\, \kappa = 1,\,m_{\rm sp}= 1,\,\lambda
= 1,\,V_0 = 1,\,V_c = 10$ and $V(0) = 0.9$. It can be noted that the
parameter $m$ enters into the equation \eqref{Vdefv} as $m^2$, hence
the sign of $m$ has no significance. As for as $n$ is concerned, we
have considered two different values of $n$, namely $n = 1/4$ that
makes the nonlinear term predominant at large $t$ and $n = 2$ when
the terms with $m$ prevails. As we have shown in our asymptotic
analysis, in both cases $q < 0$, i.e., the model expands with
acceleration. In Figs. \ref{Vblue025red2} and \ref{qblue025red2} the
blue dotted line stands for the value of $n = 1/4$, whereas the
solid red line corresponds to $n = 2$.

\begin{figure}[ht]
\centering
\includegraphics[height=70mm]{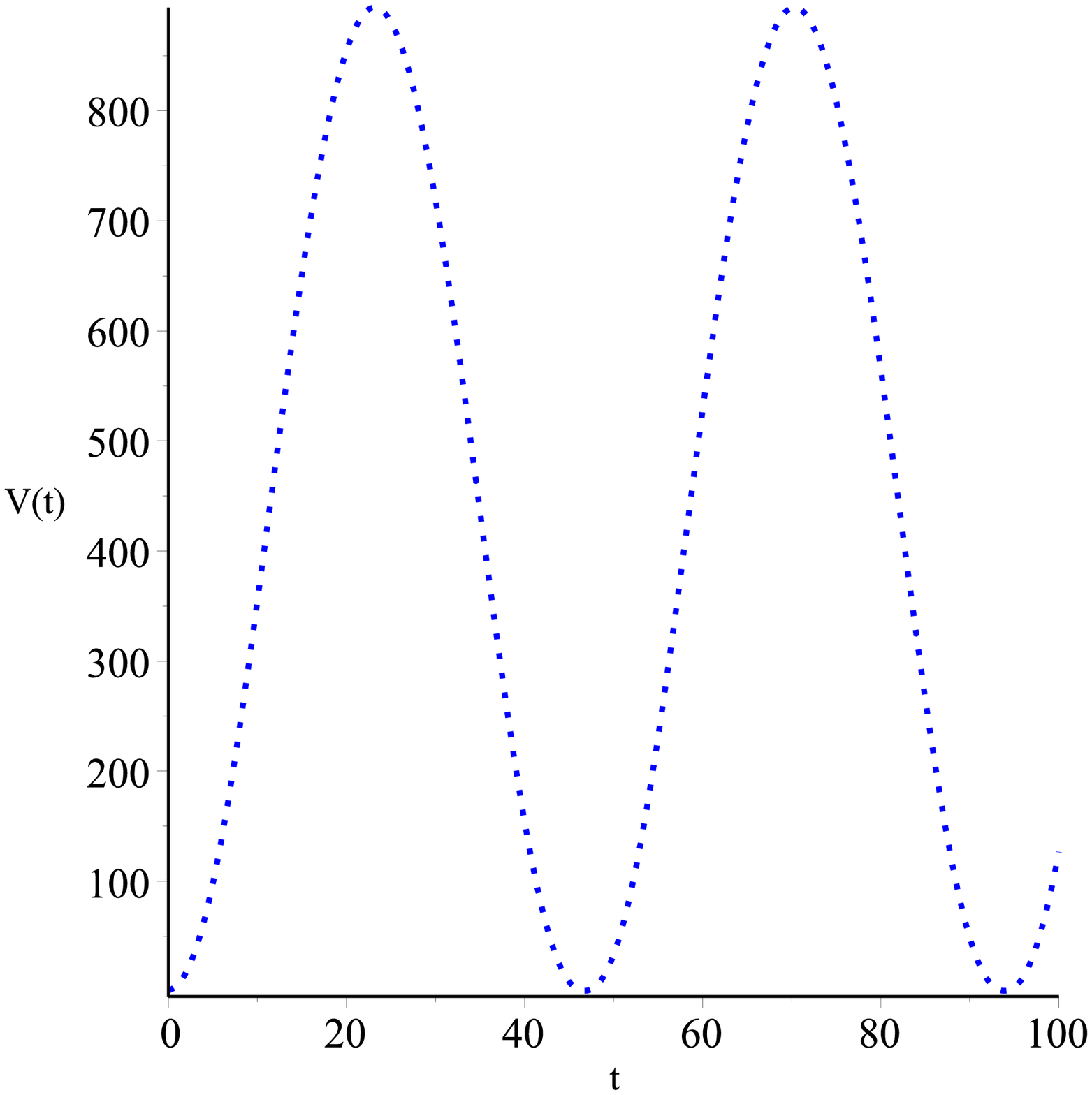} \\
\vskip 1 cm \caption{Evolution of the volume scale $V$ of the
Universe filled with massive spinor field. Here we consider the
following values of problem parameters $m =1,\,X_2 = 1,\, \kappa =
1,\,m_{\rm sp}= 1,\,\lambda = - 1,\,V_0 = 1,\,V_c = 10,\, n = 1/4$,
and $V(0) = 0.9$. The negative value of $\lambda$ gives rise to the
oscillatory mode of expansion.} \label{VoscLam1negn025}.
\end{figure}

It should be noted that in this case it is possible to obtain a
cyclic or periodical solution for $V$. In fact a negative
self-coupling constant $\lambda$ can give rise to such a solution.
In Fig. \ref{VoscLam1negn025} we have illustrated the evolution on
$V$ for a negative $\lambda$. As one sees, in this case the Universe
begin to expand from the given initial value, then attains some
maximum before contracting again.

{\bf case II} On the other hand exploiting the restriction
\eqref{A0A3b}, i.e., $A^3 = 0$ from \eqref{A03} one finds $A^0 = 0$.
Hence in this case we find $A^\mu = 0$, which, thanks to the
identity $I_A = - (I + J) = -(S^2 + P^2) = 0$ ultimately leads to $S
= 0$ and $P = 0$, i.e., in this case the spinor field nonlinearity
vanishes \cite{sahaAPSS2015}.

In this case for $V$ we find
\begin{equation}
\ddot V = 0,  \label{Vlin}
\end{equation}
with the solution
\begin{equation}
V = b_0 t + b_1, \quad b_0,\, b_1 - {\rm consts.} \label{Vlinsol}
\end{equation}
As far as spinor field is concerned the Matrix $A$ in \eqref{phi} in
this case become trivial and the components of the spinor field can
be written as
\begin{equation}
\psi_i = \frac{c_i}{\sqrt{V}}, \quad i = 1,\,2,\,3,\,4,
\label{psicomp}
\end{equation}
with $c_i$'s being the constant of integration obeying

\begin{subequations}
\begin{eqnarray}
c_1^* c_1 + c_2^* c_2 - c_3^* c_3 -c_4^* c_4 &=& 0, \label{S0new10} \\
c_1^* c_3 + c_2^* c_4 - c_3^* c_1 - c_4^* c_2 &=& 0. \label{P0new10}
\end{eqnarray}
\end{subequations}
It should be noted this result coincides with the one obtained for
Bianchi type $VI_0$ spacetime \cite{sahabvi0}. Analogical result was
also obtained for Bianchi type I spacetime, when the restriction was
imposed on the spinor field only \cite{sahaAPSS2015}.

\section{Conclusion}

Within the scope of Bianchi type-V spacetime we study the role of
spinor field on the evolution of the Universe. It is found that the
presence of nontrivial non-diagonal components of the
energy-momentum tensor of the spinor field imposes some severe
restriction to the system. In one case we found that the metric
functions are similar to each other, i.e., $a_1 \sim a_2 \sim a_3$.
Analogical result was found for a Bianchi type-I model, which as a
result evolves into a FRW spacetime
\cite{sahaIJTP2014,sahaAPSS2015}. In this case the Universe expands
with acceleration if the self-coupling constant $\lambda$ is taken
to be a positive one, whereas a negative $\lambda$ gives rise to a
cyclic or periodic solution. In the second case the spinor mass and
the spinor field nonlinearity vanish and the Universe expands
linearly in time. This results is similar to the one obtained in
\cite{sahaIJTP2014,sahaAPSS2015} for Bianchi type-I model and
\cite{sahabvi0} for Bianchi type-$VI_0$ model.

\vskip 0.4 cm

\noindent {\bf Acknowledgments}\\
This work is supported in part by a joint Romanian-LIT, JINR, Dubna
Research Project, theme no. 05-6-1119-2014/2016.

Taking the opportunity I would like to thank the referee for some
very useful suggestions those guides me to significantly improve the
manuscript.


\newpage

\begin{thebibliography}{99}



\bibitem{henneaux} M. Henneaux, Phys. Rev. D {\bf 21}, 857 (1980)

\bibitem{ochs} U. Ochs and M. Sorg, Int. J. Theor. Phys. {\bf 32}, 1531 (1993)

\bibitem{saha1997a} B. Saha and G.N. Shikin,  Gen. Relat. Grav. {\bf 29}, 1099 (1997)

\bibitem{saha1997b} B. Saha and G.N. Shikin,  J. Math. Phys. {\bf 38}, 5305 (1997)

\bibitem{saha2001a} B. Saha, Phys. Rev. D {\bf 64}, 123501 (2001)

\bibitem{saha2004a} B. Saha and T. Boyadjiev,  Phys. Rev. D {\bf
69}, 124010 (2004)

\bibitem{saha2004b} B. Saha, Phys. Rev. D {\bf 69}, 124006 (2004)

\bibitem{saha2006c} B. Saha, Phys. Particle. Nuclei. {\bf 37}. Suppl. 1,
S13 (2006)


\bibitem{saha2006e} B. Saha, Grav. $\&$ Cosmol. {\bf 12}(2-3)(46-47), 215
(2006)

\bibitem{saha2007} B. Saha,  Romanian Rep. Phys. {\bf 59}, 649 (2007).

\bibitem{saha2006d} B. Saha, Phys. Rev. D {\bf 74}, 124030 (2006)

\bibitem{greene} C. Armend$\acute a$riz-Pic$\acute o$n and P.B. Greene, Gen.
Relat. Grav. {\bf 35}, 1637 (2003)

\bibitem{ribas} M.O. Ribas, F.P. Devecchi, and G.M. Kremer, Phys. Rev. D {\bf 72},
123502 (2005)


\bibitem{souza} R.C de Souza and G.M. Kremer, Class. Quantum Grav. {\bf 25}, 225006
(2008)

\bibitem{kremer} G.M. Kremer and R.C de Souza, arXiv:1301.5163v1
[gr-qc] (2013)


\bibitem{PopPLB} N. J. Pop{\l}awski, Phys. Lett. B {\bf  690}, 73 (2010)

\bibitem{FabIJTP} L. Fabbri, Int. J. Theor. Phys. {\bf 52}, 634 (2013)

\bibitem{FabGRG} L. Fabbri, Gen. Relativ. Gravit. {\bf 43}, 1607 (2011)

\bibitem{ELKO}  L. Fabbri, Phys. Rev. D {\bf 85}, 0475024 (2012)

\bibitem{FabJMP} S. Vignolo, L. Fabbri, and R. Cianci, J. Math. Phys. {\bf 52}, 112502
(2011)

\bibitem{PopPRD} N. J. Pop{\l}awski, Phys. Rev. D {\bf 85}, 107502 (2012)

\bibitem{PopGREG} N. J. Pop{\l}awski, Gen. Releat. Grav. {\bf 44}, 1007 (2012)

\bibitem{riess} A.G. Riess {\it et al.},  Astron. J.  {\bf 116}, 1009 (1998)

\bibitem{perlmutter} S Perlmutter {\it et al.}, Astrophys. J.  {\bf 517}, 565 (1999)


\bibitem{krechet} V.G.Krechet, M.L. Fel'chenkov, and G.N. Shikin, Grav. $\&$
Cosmol. {\bf 14} No 3(55), 292 (2008)

\bibitem{saha2010a} B. Saha, Cent. Euro. J. Phys. {\bf 8}, 920 (2010)


\bibitem{saha2010b} B. Saha, Romanian Rep. Phys. {\bf 62}, 209 (2010)


\bibitem{saha2011} B. Saha, Astrophys. Space Sci. {\bf 331}, 243 (2011)

\bibitem{saha2012} B. Saha, Int. J. Theor. Phys. {\bf 51}, 1812 (2012)

\bibitem{sahaIJTP2014} B. Saha, Int. J. Theor. Phys. {\bf 53}, 1109 (2014)

\bibitem{sahaAPSS2015} B. Saha, Astrophys. Space Sci. {\bf 357}, 28 (2015)

\bibitem{sahabvi0} B. Saha, The European Physical Journal Plus {\bf 130} 208
(2015)

\bibitem{GCBVI2010} B. Saha, Gravitation $\&$ Cosmology {\bf 16}, 160 (2010).


\bibitem{svBVI} Saha B. and Visinescu M., Romainan J. Phys. {\bf 55}, 1064 (2010).

\bibitem {Hoogen}
J. Ib$\acute{a}\tilde{n}$ez, R.J. van der Hoogen, and A.A. Coley,
Phys. Rev. D {\bf 51}, 928 (1995).

\bibitem {Socorro}
J. Socorro and E.R. Medina, Phys. Rev. D {\bf 61}, 087702 (2000).

\bibitem{Weaver1} M. Weaver, Classical Quant. Grav. {\bf 17}, 421 (2000).

\bibitem{sahabv} B. Saha, Int. J. Theor. Phys. {\bf 52}, 1314 (2013).

\bibitem{sahaECAADE} B. Saha,  Phys. Part. Nuclei {\bf 45}, 349 (2014)

\bibitem{Yadav2011}
A.K. Yadav, Astrophys. Space Sci.  {\bf 335}, 565 (2011).

\bibitem{kibble} T.W.B. Kibble, J. Math. Phys. {\bf 2}, 212 (1961)

\end{thebibliography}
\end{document}